\documentstyle [epsf,12pt]{article}

\newcommand{\be}{\begin{equation}}
\newcommand{\ee}{\end{equation}}
\newcommand{\ben}{\begin{displaymath}}
\newcommand{\een}{\end{displaymath}}
\newcommand{\bea}{\begin{eqnarray}}
\newcommand{\eea}{\end{eqnarray}}

\newcommand{\A}{\alpha}
\newcommand{\B}{\beta}

\newcommand{\p}{{\bf p}}

\newcommand{\tb}[1]{\,\,\tilde{\!\!\bar{{#1}}}}

\newcommand{\bfp}{{\bf p}}

\newcommand{\pid}{$\pi d$}
\newcommand{\piN}{$\pi N$}
\newcommand{\piNNpiNN}{$\pi N\! N\!\rightarrow\!\pi N\!N$}
\newcommand{\piNpiN}{$\pi N\!\rightarrow\!\pi N$}
\newcommand{\NNtopiNN}{$N\!N\!\rightarrow\!\pi N\!N$}
\newcommand{\NN}{$N\!N$}
\newcommand{\NNpiNN}{$N\!N\!-\pi N\!N$}

\newcommand{\piNN}{$\pi N\!N$}
\newcommand{\pidpid}{$\pi d\!\rightarrow\!\pi d$}
\newcommand{\pppid}{$pp\rightarrow\!\pi^+ d$}
\newcommand{\NNpid}{$N\!N\!\rightarrow\!\pi d$}
\newcommand{\NNNN}{$N\!N\!\rightarrow\!N\!N$}

\newcommand{\pidNN}{$\pi d\!\rightarrow\!N\!N$}
\newcommand{\eqn}[1]{\label{#1}}
\newcommand{\eq}[1]{Eq.\ (\ref{#1})}

\newcommand{\fign}[1]{\label{#1}}
\newcommand{\fig}[1]{Fig.\ \ref{#1}}
\newcommand{\figs}[1]{Figs.\ \ref{#1}}

\def\@cite#1{${}^{\mbox{\small #1}}$}
\def\ctt#1#2{\cite{#1}${} {\mbox{\small -}}$\cite{#2}}

\newcommand{\bi}[1]{\bibitem{#1}}
\begin{document}
\begin{center}
      FEW-BODY DESCRIPTIONS OF THE \piNN\ SYSTEM IN THREE AND
      FOUR DIMENSIONS
\footnote{Invited talk, Sixth International Symposium on Meson-Nucleon Physics
and the Structure of the Nucleon Blaubeuren/Tuebingen, Germany, 10-14 July
1995.}
\end{center}
\bigskip

\begin{center}
      B.\ BLANKLEIDER and A.\ N.\ KVINIKHIDZE$^\dagger$ \\
{\it Physics Department, Flinders University, Bedford Park, } \\
{\it S.A.\ 5042, Australia}
\end{center}

\begin{abstract}
We summarise the recent theoretical progress in few-body descriptions of the
\piNN\ system. Previous descriptions, both three- and four-dimensional, are
shown to possess serious theoretical inconsistencies. We illustrate how
three-dimensional approaches suffer from renormalisation problems, and how
four-dimensional descriptions contain both overcounting and undercounting of
diagrams.  We then show how such theoretical problems have been recently
overcome, leading to new practical few-body equations for the \piNN\ system.
\end{abstract}

\noindent{\bf 1.\hspace{3mm} Introduction}
\bigskip

In the absence of three-body forces, Faddeev equations provide the
theoretically
exact way to describe three-body systems. It is this fact that has enabled
models of the three-nucleon system where the effect of the missing three-body
forces can be accurately studied. It may at first seem that the system
consisting of one pion and two nucleons can be similarly described in an
accurate way with Faddeev equations. This, however, is not the case. The
problem
of course, is that a pion can get absorbed by one nucleon and then emitted by
the other nucleon, thus making the standard three-body description
inappropriate. Indeed any number of pions can get created and absorbed by
nucleons, and it is clear that field theory must be used to describe the \piNN\
system, rather than the standard three-body theory of quantum mechanics. The
problem of formulating a few-body description for the \piNN\ system, that is
the
analogue of the Faddeev description for three nucleons, has by now a long
history.\cite{piNNSys} Yet until very recently, both three- and
four-dimensional
formulations have had serious theoretical inconsistencies. Here we would like
to
summarise these inconsistencies, and to describe the recently developed
few-body
descriptions that appear to overcome all these theoretical problems. In
particular, we present the new \piNN\ equations in a form that is easily
compared with previous works, while at the same time being especially
convenient
for numerical solution.

\bigskip

\noindent{\bf 2.\hspace{3mm} Three-dimensional formulation}
\bigskip


Quantum field theory requires the use of four dimensions to assure manifestly
covariant descriptions.  However, solving four-dimensional equations presents
an
enormous numerical task, and consequently three-dimensional formulations of the
\piNN\ system are most desirable and indeed have been the most numerous.

Until recently, perhaps the most sophisticated few-body description of the
\piNN\ system has been the ``unitary \NNpiNN\ model'' \ctt{Avishai}{AB_80}.
This
is a field-theoretic model based on time-ordered perturbation theory, it takes
into account pion absorption, and has the desirable properties of two- and
three-body unitarity. The essential feature of the model is that it describes
all the processes $\pi d\rightarrow \pi d$, $\pi d\rightarrow \pi N\!N$, $N\!N
\rightarrow \pi N\!N$, $N\!N \rightarrow \pi d$, and $N\!N \rightarrow N\!N$
all
within the one set of coupled equations.

The derivation of the unitary \NNpiNN\ equations is based upon truncating
Hilbert space to states of at most one pion. In practise, this means retaining
all diagrams contributing to subsystem \piN\ and \NN\ potentials, but
neglecting
all other diagrams having two or more pions in an intermediate state. Many
calculations have been performed with the \NNpiNN\ model \ctt{AB_81}{Mizutani}.
In general, one can say that the model can account for an extensive amount of
data, albeit only in a qualitative way.
\bigskip

\noindent{\em 2.1\hspace{2mm} The renormalisation problem}
\bigskip
\begin{figure}[b]
\hspace{1cm} \epsffile{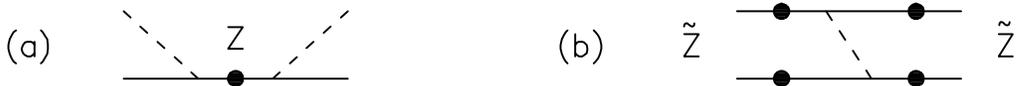}
\caption{\fign{flaw} Allowed dressing in the unitary \NNpiNN\ model, with
associated $Z$ renormalisation factors. (a) \piN\ nucleon pole graph, (b) \NN\
OPE graph.} \end{figure}

Despite the modest successes of the unitary \NNpiNN\ model, it has become clear
that the model itself has a serious theoretical
inconsistency \cite{Sauer,FewBody_Adelaide}. The origin of the problem lies in
the truncation of Hilbert space used to derive the \NNpiNN\ equations. This
truncation has serious consequences for the renormalisation of both the
two-nucleon propagator and the \piNN\ vertex.  In \fig{flaw}(a) we show the
\piN\ nucleon pole diagram where the intermediate state nucleon is dressed by
one-pion loops; however, the initial and final state nucleons do not include
dressing since two-pion states are neglected in the truncation. Since close to
the nucleon pole the dressed one-nucleon propagator is of the form $g(E) \sim
Z/(E-m)$, where $Z$ is the residue at the pole, \fig{flaw}(a) illustrates how
each \piNN\ vertex $f(E)$ gets effectively renormalised by a factor of
$Z^{1/2}$. Thus $f_{\pi N\!N} = Z^{1/2}f(m)$ is essentially the \piNN\ coupling
constant, and this fact is used to fix the strength parameter in the form
factor
$f(E)$.  With all other parameters of $f(E)$ fixed to reproduce experimental
\piN\ phase shifts, this form factor then enters the unitary \NNpiNN\ equations
as an input. As illustrated in \fig{flaw}(b), when the \NN\ one pion exchange
(OPE) amplitude is calculated in the unitary \NNpiNN\ model, the initial and
final nucleons are dressed by pions and consequently each external nucleon
obtains a renormalisation factor of $\tilde{Z}^{1/2}$. The first
renormalisation
problem is the fact that $\tilde{Z}\ne Z$. This arises because two nucleons
cannot be dressed at the same time in the truncated Hilbert space; thus, each
nucleon in a two-nucleon state cannot obtain its full dressing. This, however,
may not be such a serious problem since, in practice, the difference between
$Z$
and $\tilde{Z}$ turns out to be quite small. The serious problem, instead, is
the size of the effective \piNN\ coupling constant in the \NNpiNN\
equations. Taking $Z\approx \tilde{Z}$, \fig{flaw}(b) illustrates that each
vertex gets renormalised by a factor of $Z$, so that the effective \piNN\
coupling constant here becomes $Zf(m)$; this is a factor $Z^{1/2}$ times the
physical coupling constant. With $Z$ being typically between $0.6$ and $0.8$,
we
come to the disturbing conclusion that the effective \piNN\ coupling constant
in
the \NNpiNN\ equations is smaller than the one used in constructing the \piN\
input. This observation helps explain why one typically obtains much too small
\pppid\ cross sections using this model \ctt{Rinat2}{Fayard}.

\newpage

\noindent{\em 2.2\hspace{2mm} The \piNN\ convolution equations}
\bigskip

Here we describe a new formulation of the \piNN\ problem where unitary
equations
are obtained without having to truncate the Hilbert space to some maximum
number
of pions. Consequently, all possible dressings of one-particle propagators and
vertices are retained in our model. The essential technique that enables the
calculation of all such dressings is a novel use of convolution integrals.
In this way we overcome the renormalisation problems discussed above.

As an explicit derivation of the new \piNN\ equations can be found in
Ref.\cite{KB_PL}, here we prefer to simply state the final equations and to
describe their essential features. The new \piNN\ equations can be expressed in
many different forms, all of which are equivalent. The form we shall choose
here
is the one that most closely resembles the unitary \NNpiNN\ equations as given
by Afnan and Blankleider\cite{AB_80} (AB). Choosing this AB form has two
essential advantages: firstly, we are able to directly compare the differences
between our \piNN\ convolution equations and the unitary \NNpiNN\ equations,
and
secondly, this form is ideal for numerical solution, especially since advantage
may be taken of existing codes for the unitary \NNpiNN\ equations where
essentially the AB form has been used. We note, however, that for the sake of
easy comparison, we give the equations here only for the case of
distinguishable
nucleons. Including proper antisymmetry of the nucleons is a technical
formality
which, for the $AB$ form, will be presented elsewhere (in this regard, note
that
the derivation of the four-dimensional \piNN\ equations presented in Ref.\
\cite{KB_NP}, assumes identical nucleons from the beginning).

The \piNN\ convolution equations may be expressed as a set of coupled equations
for the reactions \NNNN, \NNpid, \pidNN, and \pidpid\ using the following
($4\times 4$) matrix form:
\be \left(
\begin{array}{cc} T_{N\!N} & \bar{\underline{T}}_{N} \\
\underline{T}_{N} & \underline{T} \end{array} \right)= \left( \begin{array}{cc}
V_{N\!N} & \bar{\underline{F}} \\
\underline{F} & G_0^{-1}\underline{\cal{I}} \end{array} \right)\left\{ I+
\left(
\begin{array}{cc} G_{N\!N} & \underline{0} \\
\underline{0} & G_0 \underline{w}^0 G_0 \end{array} \right) \left(
\begin{array}{cc} T_{N\!N} & \bar{\underline{T}}_{N} \\
\underline{T}_{N} & \underline{T} \end{array} \right) \right\}.  \eqn{AB}
\ee
Before explaining the symbols in this equation, let us define what we mean by a
product of two symbols. For any two quantities $B$ and $A$, describing
processes
$m\rightarrow k$ and $k\rightarrow n$, respectively, we define the product
symbol $AB$ to mean the the integral
\be
AB \equiv \int d\p''_1\ldots d\p''_k \,
A(\p'_1\ldots\p'_n,\p''_1\ldots\p''_k;E)
\, B(\p''_1\ldots\p''_k,\p_1\ldots\p_m;E)  \eqn{product}
\ee
where $\p_i$ is the three-momentum of particle $i$ and $E$ is the total energy.
Although, momentum conserving $\delta$-functions are assumed to be contained in
both $A$ and $B$, it is easy to see that such $\delta$-functions can be
factored
out without affecting the symbolic equations. In \eq{AB} the unknown quantities
are $T_{N\!N}$, together with $T_{\A N}$, $T_{N \B}$, and $T_{\A\B}$ which are
elements of the matrices $\underline{T}_N$, $\bar{\underline{T}}_{N}$, and
$\underline{T}$, respectively (here indices $\A$ and $\B$ take on values $1$,
$2$, and $3$). The physical amplitudes for \NNNN, \NNpid, \pidNN, and \pidpid,
are then given by
\bea
\begin{array}{ccccccc}
X_{N\!N} = T_{N\!N}  & ; &
X_{dN} = \bar{\Psi}_dT_{3N} & ; &
X_{Nd} = T_{N3}\Psi_d & ; &
X_{dd} = \bar{\Psi}_d T_{33}\Psi_d ,
\end{array}
\eea
respectively, where $\Psi_d$ is the deuteron wave function in the presence of a
spectator pion. On the r.h.s.\ of \eq{AB} $G_0$ is the fully dressed \piNN\
propagator, $G_{N\!N}$ is the fully dressed \NN\ propagator,
$\underline{\cal{I}}$ is a $3\times 3$ matrix whose elements are
$\bar{\delta}_{\A\B}=1-\delta_{\A\B}$, and $V_{N\!N}$ is the dressed one-pion
exchange potential given by
$
V_{NN}=\sum_{i,j=1}^{2}\bar{F}_i\bar{\delta}_{ij}G_0 F_j
$
where $F_i$ and $\bar{F}_i$ are fully dressed \piNN\ vertices in the
two-nucleon
sector as illustrated in \fig{F}. Finally we have the matrices
\vspace{-5mm}
\bea
\begin{array}{ccccc}
\underline{F}=\left( \sum_{j=1}^2 \bar{\delta}_{\A j}F_j \right) & ; &
\bar{\underline{F}}=\left( \sum_{i=1}^2\bar{F}_i \bar{\delta}_{i\B} \right)& ;
&
\underline{w}^0 =\left( \begin{array} {ccc}
w^0_1 & w^0_4 & 0 \\
w^0_5 & w^0_2 & 0 \\
0   & 0   & w^0_3 \end{array}  \right)
\end{array}
\eea
where the $w^0_\A$ ($\A=1\ldots 5$) are the disconnected \NN-irreducible
amplitudes for \piNNpiNN, to be discussed shortly.
\begin{figure}[t]
\hspace{3cm} \epsffile{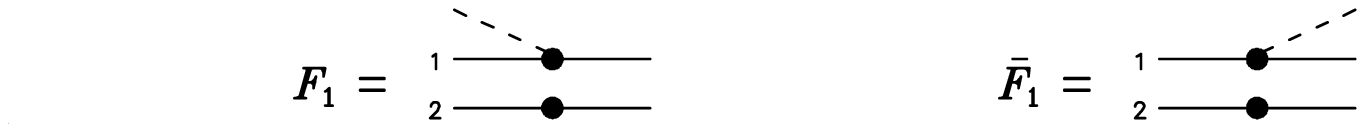}
\caption{\fign{F} The vertices $F_1$ and $\bar{F}_1$. The dark circles
represent all possible intermediate states consistent with the requirement that
$F_1$ and $\bar{F}_1$ be amplitudes with chopped external legs. Vertices
$F_2$ and $\bar{F}_2$ are obtained by interchanging 1 and 2.} \end{figure}
\begin{figure}[b]
\hspace{1cm} \epsffile{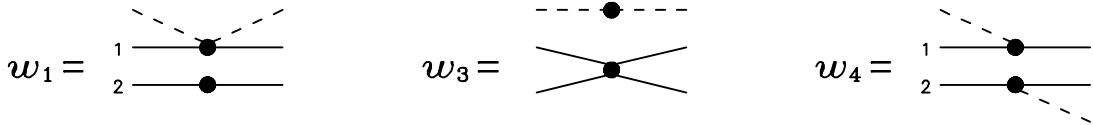}
\caption{\fign{w} The amplitudes $w_{\A}$. The dark circles represent all
possible intermediate states consistent with the requirement that the
$w_{\A}$ be amplitudes with chopped external legs. Amplitudes $w_2$ and
$w_5$ are obtained by interchanging the two nucleons in $w_1$ and $w_4$
respectively.}
\end{figure}

By form, \eq{AB} is very similar to the unitary \NNpiNN\ equations as given in
Eq.\ (59) of AB. However, the essential feature of \eq{AB} that distinguishes
it
from the unitary \NNpiNN\ equations, is that all input quantities in \eq{AB}
are
fully dressed. In this way the renormalisation problems of the \NNpiNN\
equations have been overcome. However this would only be a formal solution to
the renormalisation problem if it were not for the fact that all the necessary
dressings can be calculated exactly using convolution integrals. That this is
so
follows from Ref.\cite{KB_PRC} where we showed that any disconnected Green
function is equal to the convolution of all its disconnected parts; thus for
example, the dressed two-nucleon propagator $G_{N\!N}$ is expressed in terms of
the dressed one-nucleon propagators $g_1$ and $g_2$ as
\be
G_{N\!N}(E) = -\frac{1}{2\pi i}\int_{-\infty}^{\infty}dz\,g_1(E-z)g_2(z)
\eqn{G}
\ee
where, for the sake of simplicity, we have set the momenta of the nucleons to
zero. To further save on notation, we introduce the shorthand
$G_{N\!N}=g_1\otimes g_2$ to mean the convolution integral of \eq{G}. Giving
labels 1 and 2 to the two nucleons, and label 3 to the pion, in the same way we
have that the fully dressed \piNN\ propagator $G_0$ is given by the double
convolution
$
G_0 = g_1 \otimes g_2 \otimes g_3 .
$

To see how the amplitudes $w_\A^0$ are calculated, we first define the
amplitudes $w_\A$ to be the disconnected \piNNpiNN\ amplitudes, illustrated in
\fig{w}, each corresponding to a different type of disconnectedness, and
containing {\em all} possible contributing diagrams. It is just because the
$w_\A$ contain all possible contributions that one can express them through
convolution integrals as
\bea
\begin{array}{ccccccccc}
\tilde{w}_1 = \tilde{t}_1 \otimes g_2 & ; &
\tilde{w}_2 = \tilde{t}_2 \otimes g_1 & ; &
\tilde{w}_3 = \tilde{t}_3 \otimes g_3 & ; &
\tilde{w}_4 = \tilde{f}_1 \otimes \tb{f}_2 & ; &
\tilde{w}_5 = \tilde{f}_2 \otimes \tb{f}_1    \eqn{w_convolutions}
\end{array}
\eea
where the ``tilde'' denotes a Green function quantity consisting of the
corresponding amplitude with additional initial and final-state propagators;
thus, for example, $\tilde{w}_\A=G_0w_\A G_0$, and $\tilde{t}_1 = g_{\pi N_1}
t_1 g_{\pi N_1}$ where $t_1$ is the t-matrix and $g_{\pi N_1}$ the dressed
propagator for scattering of a pion off nucleon 1. As we have shown in
Ref.\cite{KB_PRC}, the convolution integrals effectively sum over all the
relative time orderings of one subamplitude of a disconnected diagram with
respect to another. Similarly, the vertices $F_1$ and $F_2$ are also
expressed via convolutions as
\be
\tilde{F}_1 = \tilde{f}_1\otimes g_2\hspace{1cm};\hspace{1cm}
\tilde{F}_2 = \tilde{f}_2\otimes g_1.
\ee
Once the $w_\A$ are calculated, we may then write them as
$
w_\A = w_\A^0 + w_\A^P      \eqn{w_A}
$
where $w_\A^P$ is the part of $w_\A$ that is two-nucleon reducible, while
$w_i^0$ is two-nucleon irreducible. Since we
consider all possible contributions, it is clear that ($i=1,2$)
$w_i^P = F_i G_{N\!N} \bar{F}_i$, $w_3^P = 0$, $w_4^P = F_1 G_{N\!N}
\bar{F}_2$,
and $w_5^P = F_2 G_{N\!N} \bar{F}_1.$
In this way, all the essential input to \eq{AB} has been specified.

We may finally note a second major difference between \eq{AB} and the unitary
\NNpiNN\ equations. The input matrix $\underline{w}^0$ in \eq{AB} has
off-diagonal elements, while the corresponding matrix for the \NNpiNN\
equations
is diagonal. Recalling that the amplitudes of $\underline{w}^0$ are two-nucleon
irreducible, we can see from \fig{w}, that the off-diagonal elements
$\underline{w}_4^0$ and $\underline{w}_5^0$ correspond to what has been called
the Jennings terms. As pointed out by Jennings\cite{Jennings}, these terms may
be important for the understanding of \pid\ scattering. In our case, the
Jennings terms are also fully dressed, and form an essential part of the
convolution equations. Indeed, since our \NN\ propagator $G_{N\!N}$ is fully
dressed, it also contains two-pion states coming from intermediate
Jennings-like
terms.  It is then necessary to retain $\underline{w}_4^0$ and
$\underline{w}_5^0$ in the convolution equations because they combine with
$G_{N\!N}$ in just the right way to guarantee three-body unitarity.

We recall, that the only approximation made in deriving the convolution
equations of \eq{AB} is the neglect of all connected \piNN-irreducible diagrams
for the \piNNpiNN\ process\cite{KB_PL}. Yet it is very easy to include some
types of connected contributions.  One such contribution would involve
intermediate state potentials $V_{N\!N}^{(1)}$ that are \piNN-irreducible. Then
\eq{AB} would be modified simply by replacing $V_{N\!N}$ with
$V_{N\!N}+V_{N\!N}^{(1)}$. This observation suggests that a way to include
heavy
meson exchange into our \NN\ potential would be as a phenomenological model for
$V_{N\!N}^{(1)}$.
\bigskip

\noindent{\bf 3.\hspace{3mm} Four-dimensional formulation}
\bigskip

Although three-dimensional equations may be easier to solve than those in
four-dimensions, there are important reasons why the formulation of
four-dimensional equations is necessary. Firstly, such equations are based on
relativistic quantum field theory, and retain the fundamental property of
off-shell covariance. Secondly, having the correct four-dimensional equations,
one can then do a three-dimensional reduction using one of the well-known
reduction schemes. We may also add, that with the ever increasing power of
computers, the numerical solution of four-dimensional equations becomes ever
more feasible.

The first attempts to formulate few-body equations using relativistic quantum
field theory were made already in the early 1960's \ctt{Taylor}{Tucciarone}.
Both
such general formulations and ones more specific to the \piNN\ system have been
pursued until the present time \ctt{Avishai4d}{Haberzettl}. Yet as in the
three-dimensional case, all these attempts have had theoretical
inconsistencies. In particular, all previous attempts have contained either
overcounting or undercounting of Feynman diagrams.
\bigskip
\begin{figure}[t]
\hspace*{5mm} \epsffile{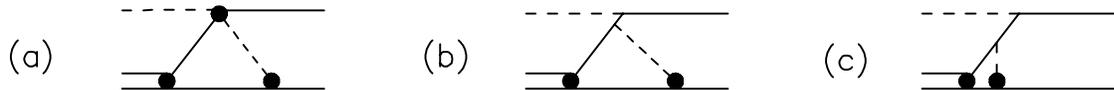}
\caption{\fign{t} Example of overcounting in \NNpid. (a) The \NNpid\
Feynman diagram where dark circles represent all possible contributions.
(b) One of the contributions included in (a). (c) Another way of drawing
diagram (b) showing how overcounting arises. } \end{figure}

\noindent{\em 3.1\hspace{2mm} Overcounting and undercounting problems}
\bigskip
\begin{figure}[b]
\hspace*{3mm} \epsfxsize=15.5cm\epsffile{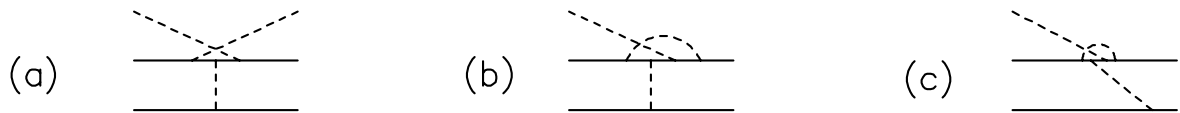}
\caption{\fign{u} Example of undercounting in \NNtopiNN. (a) A \piNNpiNN\ graph
that has usually been neglected since it corresponds to a three-body force.
(b)
The coupling of the graph in (a) to the \NN\ channel. (c) Another way of
drawing
diagram (b) reveals a two-body process. }
\end{figure}

Perhaps the easiest way to illustrate the overcounting problem in the \piNN\
system is with an example. Consider the ``triangle'' diagram of \fig{t}(a) for
the process \NNpid, where the dark circles represent the full \piNpiN\
amplitude, the dressed \piNN\ vertex, and the dressed deuteron vertex.
If one were to calculate this diagram in four dimensions, as is, using
covariant
forms for the off-shell \piN\ t-matrix, \piNN\ vertex, and the deuteron vertex,
then one would have the mistake of overcounting of diagrams. This is
illustrated
in \fig{t}(b) where we consider just the crossed-pion graph contribution to the
input \piN\ t-matrix. As these are Feynman graphs, there is no meaning
associated with the slope of the lines, and one could just as well have drawn
\fig{t}(c). However \fig{t}(c) clearly illustrates that this contribution
corresponds to the dressing of the already fully dressed deuteron vertex.

This type of overcounting arises in four-dimensional approaches whenever one
tries to formulate multiple-scattering graphs in terms of fully dressed
vertices
and full amplitudes for all subprocesses.  In once-off cases, like that of
\fig{t}(a), one can easily fix the overcounting problem by making a necessary
subtraction (here one would subtract the graph of \fig{t}(b) from the
calculation of \fig{t}(a)). However, the way to solve the overcounting problem
for the case of coupled integral equations is highly non-trivial as an infinite
number of overcounted contributions are involved.

In a similar way, let us illustrate how undercounting arises in the covariant
\piNN\ problem. As in the three-dimensional formulation, one neglects
three-body
forces also in the four-dimensional case. Only in this way can one obtain
few-body equations where (in the c.m.) no more than two independent momenta are
involved. However, one does need to be very careful about neglecting three-body
forces in the four-dimensional theory. Consider, for example the Feynman
diagram
of \fig{u}(a). This is a graph for the process \piNNpiNN\ that is both
connected
and \piNN-irreducible.  It therefore corresponds precisely to what is meant by
a
three-body force.  However, neglecting this contribution from a few-body theory
of the \piNN\ system would be a bad mistake. This is illustrated in \fig{u}(b)
where we allow the graph of \fig{u}(a) to couple to the \NN\ channel. Again no
meaning can be attached to the slope of the propagator lines, and we can
equally
well draw this diagram as in \fig{u}(c). This, however, reveals that the
three-body force of the \piNNpiNN\ process has now become a two-body
rescattering contribution in the \NNtopiNN\ process. Thus neglecting the
three-body force of \fig{u}(a) would lead to an undercounting of important
two-body contributions.
\bigskip

\noindent{\em 3.2\hspace{2mm} Four-dimensional $\pi N\!N$ equations}
\bigskip
\begin{figure}[b]
\hspace*{3mm} \epsfxsize=15.5cm\epsffile{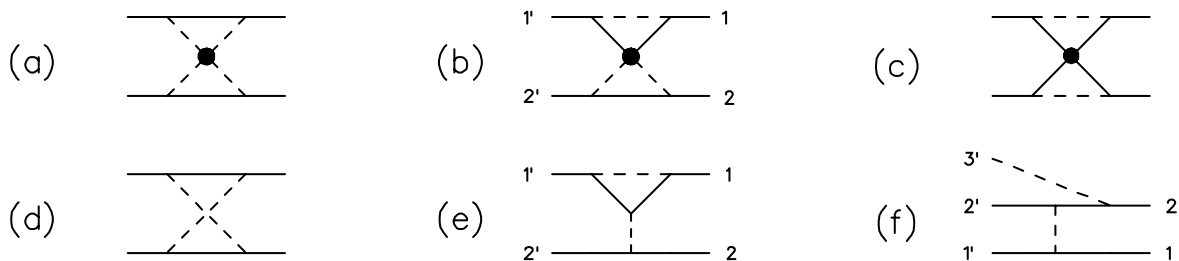}
\caption{\fign{x} The subtraction terms in the four-dimensional \piNN\
equation:
(a) $W_{\pi\pi}$, (b) $W_{\pi N}$, (c) $W_{N\!N}$, (d) $X$, (e) $Y$, and (f)
$B$. The dark circles represent the following two-body amplitudes: (a) full
$\pi\pi$ t-matrix, (b) one-nucleon irreducible $\pi N$ t-matrix, and (c) full
\NN\ t-matrix minus the \NN\ one-pion-exchange potential.}
\end{figure}

In a recent paper, we have solved both the overcounting and undercounting
problems in the formulation of few-body equations in field theory\cite{KB_NP}.
The few-body equations for the \piNN\ system then follow as a particular case.
The method used to derive the equations involves the classification of Feynman
diagrams according to their irreducibility. The overcounting problem is handled
by a procedure where, in formally identical cases like that of \figs{t}(b) and
(c), one of the two right-most vertices is ``pulled out'' further to the
right. The undercounting of diagrams is handled simply by retaining all
three-body forces until the end of the derivation where the ones that did not
lead to two-body interactions are safely neglected. It is gratifying that
Phillips and Afnan\cite{PA} have recently confirmed our equations using a
modified version of Taylor's original classification of diagram
scheme \cite{Taylor,PA2}.


With three-body forces neglected as described, one might still find it useful
to
retain, as in the three-dimensional case, the \NNNN\ \piNN-irreducible
potential
$V_{N\!N}^{(1)}$, as well as the simultaneously \NN- and \piNN-irreducible
connected \NNtopiNN\ amplitude $F_{\pi}$. However, let us at first consider
the simplest case where these contributions are neglected (they are in fact
completely absent in the usual case of a $\phi\bar{\psi}\psi$ interaction).

In this case the four-dimensional \piNN\ equations can be written as \eq{AB},
but with the following modifications: (1) The product of two quantities $A$ and
$B$ is defined as in \eq{product}, but now with all momenta and integrations
being four-dimensional (in particular the replacement $d\bfp''_i\rightarrow
d^4p_i/(2\pi)^4$ needs to be made), (2) all convolutions of Green functions are
replaced by usual products, (3) the matrix $\bar{w}^0$ is now diagonal
(i.e. with $w^0_4=w^0_5=0$), and (4) the following replacements are made,
\bea
\begin{array}{ccccc}
\underline{F}\rightarrow \underline{F}-\underline{B'} & ; &
\underline{\bar{F}}\rightarrow \underline{\bar{F}}-\underline{\bar{B}'} & ; &
V_{N\!N}\rightarrow V_{N\!N}-\Delta
\end{array}
\eea
where the terms $\underline{B'}$ and $\Delta$ are subtraction terms that
exactly compensate all the overcounting due to the use of full off-shell
amplitudes and fully dressed vertices in the coupled scattering equations. They
are defined with the help of \fig{x} as follows:
\be
\Delta = W_{\pi\pi} + W'_{\pi N} + W_{N\!N} + X + Y' - \bar{B}'G_0B'
\eqn{sub}
\ee
where the dashed quantities are the sums $W'_{\pi N}=W_{\pi N}+PW_{\pi N}P$,
$Y'=Y+PYP$, and $B'=B+PBP$, $P$ being the nucleon label exchange operator.
$\underline{B'}$ is a column matrix with each element given by $B'$.

In the more general case where $V_{N\!N}^{(1)}$ and $F_{\pi}$ are retained, it
turns out that only the subtraction terms need be modified.  In
particular, we need to do the replacements $\Delta \rightarrow \Delta-
V_{N\!N}^{(1)}-\bar{F}_\pi(F_1+F_2)+
\bar{F}_{2\pi}F_1F_2-(\bar{F}_1+\bar{F}_2)F_\pi-\bar{F}_1\bar{F}_2F_{2\pi}$ and
$B'\rightarrow B'-F_\pi$, where $F_{2\pi}$ is the connected \NN- and \piNN-
irreducible amplitude for $N\!N\rightarrow \pi\pi N\!N$.
\vspace{1cm}

\noindent
$^\dagger$ Permanent address:\,\, Mathematical Institute of Georgian Academy of
Sciences, \\ Z. Rukhadze 1, 380093 Tbilisi, Georgia.

\end{document}